\documentstyle[aas2pp4]{article} 
\tightenlines

\begin{document}
\title{Numerical Simulations of Oscillation Modes \\
of the Solar Convection Zone
}

\author{D. Georgobiani\altaffilmark{1},
A.G. Kosovichev\altaffilmark{2},
R. Nigam\altaffilmark{2,3},
A. Nordlund\altaffilmark{4},
R.F. Stein\altaffilmark{1}}

\altaffiltext{1}{Physics and Astronomy Department,
Michigan State University, East Lansing, MI 48824-1116}

\altaffiltext{2}{W.W.Hansen Experimental Physics Laboratory,
Stanford University, Stanford, CA 94305-4085}

\altaffiltext{3}{Department of Mathematics,
Stanford University, Stanford, CA 94305-2125}

\altaffiltext{4}{Teoretisk Astrofysik Center, Danmarks
Grundforskningsfond, Juliane Maries Vej 30, DK-2100
K{\o}benhavn \O, Denmark}

\begin{abstract}
We use the three-dimensional hydrodynamic code of Stein and Nordlund to
realistically simulate the
upper layers of the solar convection zone in order to study physical
characteristics of solar oscillations.  Our first
result is that the properties of oscillation modes in the simulation closely
match the observed properties.  Recent observations from SOHO/MDI and GONG
have confirmed the asymmetry of solar oscillation line profiles,
initially discovered by Duvall et al.  In this paper we compare the line
profiles in the power spectra of the Doppler velocity and continuum
intensity oscillations from the SOHO/MDI observations with the simulation.
We also compare the phase
differences between the velocity and intensity data. We have found that
the simulated line
profiles are asymmetric and have the same asymmetry reversal between
velocity and intensity as observed. The phase difference
between the velocity and intensity signals is negative at low frequencies and
jumps in the vicinity of modes as is also observed.  Thus, our numerical
model reproduces the basic observed properties of solar oscillations,
and allows us to study the physical properties which are not observed.
\end{abstract}


\section{Introduction}
The peaks of solar oscillation modes observed in velocity and intensity
power spectra are asymmetric (Duvall et al. 1993).
Moreover, the asymmetry between the velocity and intensity line profiles
is reversed.
This latter was a puzzling result, and it was initially thought to be an 
error in
the observations (Abrams \& Kumar 1996). However, recently it was
confirmed by SOHO/MDI observations (Nigam et al, 1998).
The velocity power spectrum has negative asymmetry (more power
on the low frequency side of the peak), while the
intensity power spectrum has positive asymmetry (more power
on the high frequency side of the peak).
In general, the asymmetry is a result of excitation of solar
oscillations by a localized source. It originates from the
interference of direct waves from the source with waves
that start inward and are refracted back out.
The asymmetry is a strong function of frequency,
and varies weakly with angular degree $\ell$. The reversal of asymmetry
between velocity and intensity is due to the presence
of correlated background noise, whose level depends on characteristics
of granulation
(Nigam et al, 1998). Since
the model of Nigam et al (1998) is a phenomenological one,
the physics of the correlated noise is not yet fully understood.
Roxburgh \& Vorontsov (1997) proposed that the reversal in asymmetry
occurs in velocity.  Observations suggest that the reversal occurs in
intensity as proposed by Nigam et al (1998).  It is therefore desirable
to test these ideas. 
Realistic 3D hydrodynamic simulations of upper layers of the solar
convection zone (e.g., Stein \& Nordlund, 1989; Stein \& Nordlund,
1998) have p and f modes with similar asymmetries and asymmetry reversals as the
observed modes.
These simulations can therefore be used to study the
characteristics of the correlated noise.  

Another interesting property of solar oscillation modes is the phase difference
between velocity and intensity, which was first observed by Deubner
\& Fleck (1989) and studied
theoretically by Marmolino \& Severino (1991). It may 
provide an useful diagnostic of the excitation mechanism
of the oscillations. Severino et al. (1998), Straus et al. (1998),
Oliviero et al. (1999) and Nigam \& Kosovichev (1999) attribute the
phase behavior to the interaction of the correlated background with
the oscillations. These phase relations can also be studied using
the realistic hydrodynamic simulations of the near surface layers of the Sun.

\section{Numerical Model of Convection}

We use the numerical code of Stein and Nordlund
(1998) to make a physically
realistic three-dimensional simulation of the shallow upper layers
of the solar convective zone. The code solves the
compressible hydrodynamic equations and includes ionization in the
equation of state and LTE radiative transfer in the energy balance.
The horizontal boundary conditions are periodic,
while the top and bottom boundaries of the computational domain
are transmitting. The horizontal
size of the domain is 6 by 6 Mm. It ranges in depth
from 2.5 Mm below the $\tau = 1$ surface to 0.5 Mm above it,
reaching the height of the temperature minimum of the solar atmosphere.
The computational grid of
$63 \times 63 \times 63$ mesh points gives a spatial resolution
of 100 km $\times$ 100 km horizontally and 35 -- 90 km vertically.
The system becomes thermally relaxed after several turnover times
(a turnover time is approximately 1 hour).
We have generated data for 43 hours of solar time, which provide
6.4 $\mu$Hz frequency resolution.  We calculate power
spectra of the velocity and intensity and their phase
differences to study properties of the oscillations generated in this
simulation box.
In the simulation, the
first set of nonradial modes corresponds to a harmonic degree $\ell =
740$ (or $k_h = 1$ Mm$^{-1}$), the second set corresponds to $\ell = 1480$
($k_h = 2$ Mm$^{-1}$) etc.

\section{Calculation of the Power Spectra}

Solar oscillations in velocity and intensity are measured from the Ni I
6768\AA$\;$ absorption line which is formed about 200--300 km above the
photosphere. The MDI instrument on SOHO spacecraft records filtergrams
(intensity images
at five wavelengths) which span the absorption line.  The velocity signal
is obtained by differencing the filtergrams on opposite sides of the
line which is sensitive to Doppler shift and minimizes the effect of
intensity fluctuations.  The intensity signal is obtained by summing the
filtergrams in a way to approximate the continuum intensity in the vicinity
of the spectral line and minimize the
effects of Doppler shifts (Scherrer et al. 1995).
For the comparison with the numerical simulations,
we use a time series of the spherical harmonic
transform of the full-disk data for $\ell = 740$.
The effect of solar differential rotation is removed;
and the oscillation power spectra are summed over all $2 \ell+1$ $m$ values.
We note that MDI measures the continuum intensity in the vicinity of the
Ni I 6768\AA$\;$ line,
which is different from the broad-band continuum data obtained from the
simulations. This might be a
source of some of the divergences between the model
and the observations.

In the numerical model, the vertical component of velocity, $V(x,y,t)$,
is calculated
at a height of 200 km above the $\tau = 1$ surface,
close to the formation height of the observed MDI
Doppler velocities. The simulated continuum intensity
(the emergent continuum specific intensity), $I(x,y,t)$,
is calculated by solving the Feautrier equation
along a vertical ray, in LTE, using an opacity distribution function.
To extract the nonradial modes we multiply
$V(x,y,t)$ and $I(x,y,t)$ by
$\sin(k_x x)$, $\sin(k_y y)$, $\cos(k_x x)$ and $\cos(k_y y)$,
where $k_x$ or $k_y$ are horizontal wave numbers equal to
$2 \pi n / L$,  where $L = 6$ Mm is the horizontal dimension
of the simulation box (the same for both $x$ and $y$ directions), and
$n = 1,2,3,...$ is the number of nodes in horizontal direction.
Then we  average the products of $V$ and $I$ with these sines and
cosines over the horizontal planes.
This corresponds to the 2-dimensional
Fourier transform of the
data with specific horizontal wave numbers. We then
take Fourier transforms in time to obtain the power
spectra, $\tilde{V}(k_x,k_y,\omega)$ and $\tilde{I}(k_x,k_y,\omega)$,
and sum these four different spectra to obtain a
power spectrum for a
particular horizontal wave number $k_h^2 = k_x^2 + k_y^2$.
This corresponds to summing the observational power spectra
over azimuthal degree $m$.
Since the line profiles for $k_h = 2$ Mm$^{-1}$ ($\ell=1480$)
and higher are rather noisy,
we present the results for $k_h = 1$ Mm$^{-1}$ ($\ell=740$) only.
The $k_h = 1$ Mm$^{-1}$ spectra are obtained
either for $n_x = 1$ and $n_y = 0$ or $n_x = 0$ and $n_y = 1$.
The modulus and phase of the cross spectra
$\tilde{I} \tilde{V}^*$ give
the coherence and phase difference between intensity
and velocity.

\section{Comparison between the Observed and Simulated
Power Spectra}

\begin{figure}[ht]
\epsscale{1}
\plotone{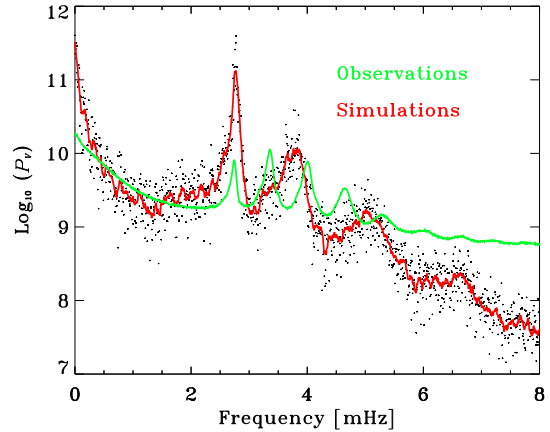}
\caption{\footnotesize Observed and simulated velocity power spectral density for $\ell =
740$ from the SOHO/MDI (green curve) and for the first non-radial mode
of a simulation of solar surface convection (dots and red curve).
Units are (cm/s)$^2$/Hz. Observed velocities are from the Doppler shifts 
of the NiI 6768 {\AA} line summed over all $2 \ell +1$ $m$ modes.
Simulated
velocities are from a calculation on a domain 6 $\times$ 6 Mm 
horizontally $\times$ 2.5 Mm deep, spanning 43 hours of solar time,
summed over the 4 modes with 6 Mm horizontal wavelength.  The simulated
modes are sparser and broader than the observed modes because the
simulated domain is shallower than the turning points of the observed
modes, resulting in a smaller mass and larger amplitude for the
simulated modes.}
\label{Vpower1}
\end{figure}

\begin{figure}[ht!]
\epsscale{1}
\plotone{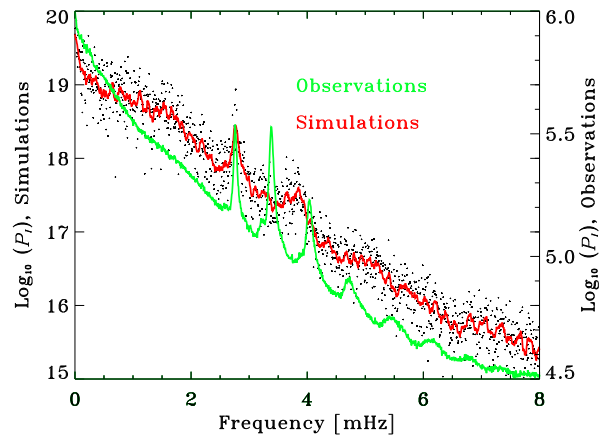}
\caption{\footnotesize Observed and simulated intensity power spectral density for the same
case as the velocity in Figure \ref{Vpower1}.  Observed intensity power
spectral density is in units of (CCD counts per sec)$^2$/Hz.  Simulated
intensity power spectral density is in units of
(erg/cm$^2$/s/ster)$^2$/Hz.
}
\label{Ipower2}
\end{figure}

In Figures \ref{Vpower1} and \ref{Ipower2} we compare the observed and
simulated
mode power spectral densities in velocity and intensity for $\ell=740$.
The observational data represent 5 days of the
observations of a quiet-sun region with a spatial resolution of 2 arcsec/pixel
($\sim 1,500$ km)
and temporal
resolution 1 min.
The simulation data are for the lowest non-radial mode of the same
angular degree,   $\ell=740$
($k_h = 1$ Mm$^{-1}$) calculated from a 43 hour run
with a horizontal grid spacing of 100 km and time spacing of 0.5 min.
In order to make our comparison more appropriate, we took only 43 hours 
of the observational data and every other snapshot in the simulated 
data to have the same 1 min time step as in observations. 
The simulated and observed velocity power are similar between the
peaks at low frequency, but the peak power  is larger in the 
simulation than the observations because the same excitation and 
damping processes are supplying energy to fewer modes with smaller
mode mass. The simulation power also falls off much more rapidly at high
frequencies, probably due to the low resolution of the simulation.
The frequency separation between the
simulated modes is larger than the separation between the observed modes
because
of the shallow computational domain.  It
is only 2.5 Mm deep, whereas $\ell=740$
modes in the Sun have the turning points below $\sim 4$ Mm.
As a result, the resonant frequencies in the simulation
are different from the frequencies of solar modes.
Also, the simulated modes
have less inertia and larger amplitudes than the solar modes.
Because of the smaller inertia (mode mass) the 
damping rate is much larger in the simulation than the Sun, so
the simulated line profiles are broader than the observed ones.
However, these differences do not prevent us from studying the
mode physics with the numerical simulations.


The lowest frequency mode in the power spectra is the f mode, which
is essentially a surface gravity mode. It does not exist
in radial oscillations, but in the nonradial data,
it is very strong and shows the same asymmetry behavior
as do p modes. Interestingly enough, in the simulations the amplitude of
the f mode  is greater than the amplitudes of the p modes, whereas in
the observational data
the f-mode amplitude is smaller than the amplitude of the p$_1$ mode.

For both the simulations and the observations the
velocity and intensity line profiles
are clearly asymmetric, with
opposite asymmetries in velocity and intensity
(Figures \ref{Vpower1} and \ref{Ipower2}).
Also, the slopes of the power spectra
in velocity and intensity have a 
power law behavior in both the simulation and observations. 

\section{Comparison between the Observed and Simulated
Intensity - Velocity Phase Difference.}

\begin{figure}[ht!]
\epsscale{1.00}
\plotone{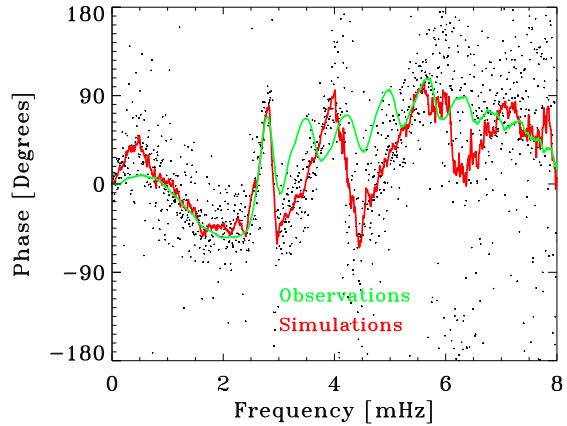}
\caption{\footnotesize Observed and simulated intensity - velocity phase
difference.  The observed data are from
SOHO/MDI with $\ell=740$ summed over all $2 \ell +1$ $m$ values.
The simulated data are for the first non-radial mode (with horizontal
wavelength 6 Mm).  The phase jumps $\sim 90^\circ$ at the mode
resonant frequencies where the
oscillation amplitude is large.}
\label{V-I-phase3}
\end{figure}

Figure \ref{V-I-phase3} compares the
observed and simulated phase differences between the intensity and
velocity signals, $I-V$, as a function of frequency.  Both the observations and
simulations have negative phase difference $I-V$ at low frequency.
In both cases, the phase difference departs from the
$90^\circ$ phase difference predicted by the adiabatic theory of
trapped standing waves.  In both cases, there are jumps in phase
of  $\sim 90^\circ$ at the
mode frequencies. The phase differences approach zero for
high frequency propagating waves as expected for adiabatic acoustic
waves.


\section{Discussion}

Mode asymmetry in the spectral domain is a result of the interference
pattern of waves from a localized source not being symmetrical relative
to the mode resonant frequencies.  Constructive and
destructive interference occurs between waves from the source that propagate
outward with those that propagate inward and are refracted outward.
Destructive interference causes characteristic ``troughs''
on one side of the modal lines, where the amplitude is minimal.
The frequencies of the ``troughs''
depend on the source location and type (Duvall et al. 1993, Vorontsov et al.
1998, Nigam and Kosovichev, 1999).
In the solar case, the destructive
interference occurs at frequencies slightly higher than the resonant
frequencies. This corresponds to the negative line asymmetry.
The interference pattern may be also affected by the
presence of an additional signal correlated with the source. 
Variations of the continuum intensity in the convective
granules, which are associated with the process of excitation
of solar oscillations, are such a signal (Nigam et al., 1998). 
This correlated signal shifts the frequencies of the ``troughs''.
If the correlated signal is sufficiently strong the
``troughs'' may be shifted to the other side of the mode
resonant frequencies, and thus reverse the asymmetry of the modal
lines.
This reversal of the asymmetry is observed in both the MDI intensity
data and the intensity spectra obtained in the numerical simulations.
The numerical results allow us to verify that  variations in the background
intensity in the
granules are indeed correlated with the oscillations (Nordlund and Stein
1998).
In the MDI Doppler velocity measurements this correlated
variation is essentially cancelled by taking differences of values on
opposite sides of the line.  In the intensity measurement it is
enhanced by summing contributions from opposite sides of the line.  The
correlated component of the noise must be large enough to reverse
the asymmetry in the intensity power spectrum, but not sufficiently large to
reverse the asymmetry in the velocity power spectrum
(Nigam et al. 1998).

These results are  consistent with
the observations of Goode et al. (1998), where it appears that the
acoustic events occur in cool and narrow intergranular dark lanes.
Prior to an acoustic event there is darkening in the intensity. This
darkening is also correlated to the strength of the event.  Strong
events are preceded by longer darkening.  The reversal of asymmetry in
intensity due to a correlated background has been confirmed by Kumar \&
Basu (1999) and Rast (1999).  However, there is, as yet, no complete
theory of the influence of the background variations on the observed
properties of solar oscillations.

The variations of the phase difference between the intensity and
velocity signals with frequency can also be explained by
 the presence of the correlated
noise (Nigam \& Kosovichev 1999).
The intensity - velocity cross spectrum is the product
\begin{displaymath}
  C_{I-V} = (\delta I + N_I)(\delta V + N_V)^* +
           N_{I {\rm uncorr}} N_{V {\rm uncorr}}\ ,
\end{displaymath}
and the intensity - velocity phase difference is
\begin{displaymath}
\tan \Theta_{I-V} = \Im C_{I-V} / \Re C_{I-V}
\ .
\end{displaymath}
Away from a mode the noise contribution dominates, while at an
eigenfrequency the mode amplitude is large and dominates.  If
the intensity and velocity are nearly $90^\circ$ out of phase, as
they are expected to be for adiabatic waves, then they will
contribute primarily to the imaginary part in the numerator, but
have little effect on the denominator.  At the mode
eigenfrequency there will be a large jump in mode amplitude,
producing a large jump in the ratio $\tan \Theta_{V-I}$, and
hence a change in $\Theta$ of order $\sim 90^\circ$.

The phase and asymmetry behavior of the modes
constrains the nature of the excitation mechanism
of solar oscillations.

\section{Conclusions}

In this Letter, we have compared observed
asymmetries of the oscillation
line profiles in the velocity and
intensity power spectra with those obtained in 3D hydrodynamic simulations
of solar convection.  The basic
characteristics of the observed mode peak asymmetries are
reproduced in the simulations: the modes are asymmetric, with
the velocity and intensity having opposite asymmetries.  The
reversal in asymmetry occurs in the intensity signal (possibly due to the
process of radiative transfer, which causes additional signal
in the intensity fluctuations correlated with the oscillation).
This is consistent with the result of Nigam et al (1998).
This is an important step in studying the
physical properties of solar oscillations and their interaction
with turbulence.  The basic characteristics of the observed
intensity -- velocity phase difference are also reproduced in
the simulations.  The similarity of the oscillation mode
properties in the simulation and observations means that the
simulations can be used to investigate the origin of mode
behavior.

\acknowledgments
This work was supported by the SOI-MDI NASA contract NAG5-3077
at Stanford University and by
NASA grant NAG 5-4031 and
NSF grant AST 9521785 to MSU.
D.G. expresses her gratitude to Sasha Kosovichev for the
possibility of performing a part of this work at Stanford, and to
Maurizio Oliviero for useful discussions on phase differences.
The calculations were performed at Michigan State University and
the National Center for Supercomputing Applications.
SOHO is a project of international cooperation between ESA and NASA.

\end{document}